\documentstyle[prd,aps]{revtex}
\begin{document}
\draft
\twocolumn[\hsize\textwidth\columnwidth\hsize\csname
@twocolumnfalse\endcsname
 \preprint{SU-ITP-96-49,  astro-ph/9610219}
\title{Nongaussian Isocurvature  Perturbations from Inflation}
\author{A. Linde$^\ast$ and V. Mukhanov$^\dagger$}
\address{$^\ast$Physics Department, Stanford University,
Stanford CA 94305-4060, USA\\
$^\dagger$Institute for Theoretical Physics, ETH, Honggerberg 8093, Zurich,
Switzerland}
\date{October 21, 1996}
\maketitle
\begin{abstract}
We present a class of very simple inflationary models of two scalar fields
which leads  to   nongaussian isothermal perturbations with   ``blue" spectrum,
$n > 1$.  One of the models is inspired by supersymmetric theories where light
scalar fields   naturally acquire masses $\sim H$ during inflation. Another
model presumes that one of the fields has a nonminimal  interaction with
gravity $\xi R \sigma^2$. By a slight modification of parameters of these
models one can obtain either gaussian isothermal perturbations, or nongaussian
adiabatic perturbations with  $n > 1$.
\end{abstract}
\pacs{PACS numbers: 98.80.Cq \hspace{2.6 cm} SU-ITP-96-49 \hspace {2.6 cm}
 astro-ph/9610219}
\vskip2pc]

\section{Introduction}
It is well known that quantum fluctuations of the inflaton field $\phi$ in the
simplest inflationary models
produce  gaussian adiabatic  perturbations of metric with nearly scale
independent spectrum, $n \approx 1$~ \cite{Pert,muk1}.   In the models
involving several scalar fields one can also obtain significant  isothermal
perturbations \cite{Ax}. However, such perturbations  produce six times greater
anisotropy of the microwave background radiation than adiabatic perturbations
(for a review see \cite{muk1,Liddle}. To obtains small MBR anisotropy and still
explain galaxy formation from isothermal perturbations one should strongly
suppress entropy perturbations on the horizon scale while keeping them
sufficiently large on the galaxy scale. This would be possible, for example, if
  entropy perturbations  had ``blue'' spectrum decreasing at large scales with
$n {\ \lower-1.2pt\vbox{\hbox{\rlap{$>$}\lower5pt\vbox{\hbox{$\sim$}}}}\ }
1.5$. However, until now there were no inflationary models which would produce
such entropy perturbations. In principle, it would be possible to obtain such
perturbations in axion models \cite{Ax}, but in the same theories where one
could have  $n {\
\lower-1.2pt\vbox{\hbox{\rlap{$>$}\lower5pt\vbox{\hbox{$\sim$}}}}\ } 1.5$ one
would also have too many axion strings with domain walls attached to them
\cite{ALyth}. Therefore typically the authors discussing possible observational
consequences of inflationary entropy perturbations concentrated on the models
with $n \leq 1$, where only a minor addition of entropy perturbation to
adiabatic ones was admissible \cite{iso}.

In this paper we will propose a class of very simple models where one can get
either entropy or adiabatic perturbations with  $n {\
\lower-1.2pt\vbox{\hbox{\rlap{$>$}\lower5pt\vbox{\hbox{$\sim$}}}}\ } 1.5$. This
possibility may be of particular interest in relation to open universe models
\cite{martin}.  The main idea  is almost trivial: If one has a theory of a
scalar field $\sigma$ with a mass $m$, then   the mode functions $\sigma _k(t)$
in the expansion of field
operator $\hat \sigma (x,t)$
\begin{equation}
\hat \sigma (x,t)=\frac 1{\sqrt{2}}\int \left[ \sigma _k^{*}(t)e^{ikx}\hat
a_k^{-}+\sigma _k(t)e^{-ikx}\hat a_k^{\dagger }\right] \frac{d^3k}{\left(
2\pi \right) ^3}  \label{x1}
\end{equation}
obey the differential equation
\begin{equation}
\ddot \sigma _k+3\frac{\dot a}a\dot \sigma _k+({k^2\over a^2}+m^2)\sigma _k=0 \
{}.
\label{x2}
\end{equation}
Here $k$ is a comoving wavenumber.  We will normalize the scale factor $a$ in
such a way that $a = 1$ at the end of inflation. At this time in the most
interesting comoving wavenumber interval $k<
H$ the mode functions $\left| \sigma _k(t)\right| $ for $m^2 \ll H^2$ become
\cite{fluct}
\begin{equation}
 |\sigma _k^2(t)| k^3 =   {H^{2}\over 2} \left({k\over H}\right) ^{2m^2\over 3
H^2},
\label{x3}
\end{equation}
which corresponds to the spectral index $n \approx 1 + {4m^2\over 3 H^2} >1$.
The estimate of energy density of   fluctuations of the field $\sigma$ can be
obtained as follows:
\begin{equation}\label{1a}
\rho_\sigma \simeq { m^2 \langle \sigma^2 \rangle\over 2}  = {H^2 m^2\over
8\pi^2} \int\limits_0^H {dk\over k}\left({k\over H}\right)^{2m^2\over 3 H^2} =
{3H^4\over 16 \pi^2} \ .
\end{equation}
The main contribution to the energy density is given by the modes with $k_0 < k
<H$, where $k_0$ is exponentially small, $k_0 \sim H \exp\bigl(-{H^2\over
m^2}\bigr)$.  Let us consider density perturbations of the field $\sigma$ in
the long wavelength limit, $k \ll k_0$. This limit is most interesting in the
models with sufficiently large $m/H$, such that the wavelengths $k_0^{-1}\sim
H^{-1} \exp\bigl({H^2\over m^2}\bigr)$ after their subsequent growth in an
expanding universe   still remain smaller than the galaxy length scale. In this
case on the scale $k^{-1} \gg k_0^{-1}$ the fluctuating field $\sigma$ wanders
many times in the region $-H^2/m {\
\lower-1.2pt\vbox{\hbox{\rlap{$<$}\lower5pt\vbox{\hbox{$\sim$}}}}\ }\sigma {\
\lower-1.2pt\vbox{\hbox{\rlap{$<$}\lower5pt\vbox{\hbox{$\sim$}}}}\ } H^2/m$, so
that its value averaged over the domain of a size $k^{-1}$ vanishes. As a
result, addition of a perturbation $\delta\sigma (k)$ with $k \ll k_0$ does not
lead to usual density perturbations $m^2\sigma\delta\sigma$ \cite{Lyth,Schmid}.
Perturbations of density will be nongaussian and quadratic in $\delta\sigma$;
they can be very roughly estimated by
\begin{equation}\label{6i}
| \delta _k^\rho | \sim {\delta\rho_\sigma(k)\over \rho_\sigma}   \sim {2\pi
m^2|\sigma _k^2(t)| k^3\over  \rho_\sigma} \sim
{m^2\over H^2 }\left({k\over H}\right)^{2m^2\over 3 H^2}  ,
\end{equation}
which implies that
\begin{equation}\label{6}
{\delta\rho_\sigma\over \rho_{\rm total}} \sim | \delta _k^\rho |
\,{\rho_\sigma\over \rho_\phi}  \sim
{m^2 \over  M_p^2}\left({k\over H}\right)^{2m^2\over 3 H^2} \ .
\end{equation}
In (almost) all realistic models of inflation the ratio ${m^2\over M_p^2}$ is
extremely small \cite{book}. Therefore initially perturbations of the field
$\sigma$ practically do not contribute to perturbations of metric (isocurvature
 perturbations).  However,   the inflaton field must decay to produce matter.
If the products of its decay are ultrarelativistic particles, then their energy
density subsequently decreases as $a^{-4}$, where $a$ is the scale factor of
the universe. Meanwhile, if the field $\sigma$ does not decay, or decay very
late, then for a long time its energy density   decreases more slowly, as
$a^{-3}$. As a result,  perturbations of the field $\sigma$  may give rise to
significant  perturbations of metric. This is the same mechanism which is
responsible for isothermal perturbations in axion theory \cite{Ax}. However,
the axion mass during inflation vanishes. In our case   $m$ does not vanish,
and  eq. (\ref{6}) implies that  the perturbations  have blue spectrum, which
is exactly what we need.

It is not very easy to implement this idea. First of all, Hubble constant can
be very large in the beginning of inflation, and during the last stages of
inflation it may change significantly. For example, in the chaotic inflation
scenario in the theory $\lambda\phi^4$ inflation begins at $H \sim M_p$, and
stops at $H \sim \sqrt\lambda M_p$. One must fine-tune the parameters   to
ensure that $H$ becomes comparable to $m$ exactly at the epoch corresponding to
the last 60 e-folds, when the perturbations responsible for the large scale
structure of the observable part of the universe have been formed. Moreover,
prior to this epoch $H$ remains much greater than $m$. But then there is no
reason to assume that the field $\sigma$ initially was at the point $\sigma =
0$: it did not have any chance to roll down there at $H \gg m$.

In this paper we will propose a simple way to overcome these problems. It is
based on the observation that masses of many scalars in supersymmetric theories
(except, possibly, the mass of the inflaton field) acquire corrections $\Delta
m^2 = \alpha H^2$ in the early universe \cite{moduli}. Here $\alpha$ is some
parameter; typically $\alpha = O(1)$, but exceptions are possible. To study
perturbations of metric which may appear in such models we will consider here a
  simplest toy model where the scalar field $\sigma$ acquires the mass $\sim H$
during inflation. It describes   fields $\phi$ and $\sigma$  with the
effective potential
\begin{equation}\label{4}
V(\phi,\sigma) =  {1\over 2}{M^2\phi^2 }+ {1\over 2}{m^2\sigma^2 } + {1\over
2}{g^2} \phi^2\sigma^2 \ .
\end{equation}

This potential has two valleys, at $\sigma = 0$ and at $\phi = 0$. If
initially $|\phi| > |\sigma |$, the field $\sigma$ rapidly rolls to the valley
$\sigma = 0$ and stays there. In this regime the energy density  during
inflation becomes equal to ${M^2\phi^2\over 2}$, and the Hubble constant is
$H^2 = {4\pi M^2\phi^2\over 3 M_p^2}$. Then the potential (\ref{4})
can be written as follows:
\begin{equation}\label{5}
V(\phi,\sigma) =  {1\over 2}{M^2\phi^2}+ {1\over 2}{ (m^2 + \alpha H^2)\sigma^2
}  \ ,
\end{equation}
where $\alpha = {3g^2 M_p^2\over 4\pi M^2}$.  Thus this simple model   can  be
considered as a toy model exhibiting generation of an effective mass squared
$\alpha H^2$ of the field $\sigma$ during inflation.

Similar mass term appears during inflation even if the fields $\phi$ and
$\sigma$ do not interact with each other ($g = 0$), but  the field $\sigma$ is
nonminimally coupled to gravity. Indeed, the term ${\xi\over 2}R\sigma^2$ does
not lead to any corrections to the rate of expansion of the universe in the
regime $\sigma = 0$, but it gives a contribution $12\xi H^2$ to the effective
mass of the field $\sigma$ during inflation \cite{SS}. In what follows it will
not be very important  which   mechanism  gives a contribution $\sim H^2$ to
the mass of the field $\sigma$; we will just assume that during inflation the
field $\sigma$ has an effective  mass squared  $m^2(H) = m^2  + \alpha H^2$.
Note that if $H$ changes slowly, then the field $\sigma$ at large $H$
approaches the state $\sigma = 0$ as follows: $\sigma = \sigma_0
\exp{\Bigl(-{m^2(H)\, t\over 3H}\Bigr)} \sim \sigma_0 \exp {\bigl(- {\alpha H\,
t\over 3}\bigr)} $. Thus, unlike in the situations where the mass is constant,
in our case the field $\sigma$ within the time $\sim (\alpha H)^{-1}$ rolls
down to $\sigma = 0$, and we may study its quantum fluctuations near this
point. Note also, that now the relation $m^2(H) = \alpha H^2$  is satisfied at
all large $H$, so that after fixing the parameter $\alpha = O(1)$ one does not
need  to fine-tune $m$ to be of the same order as $H$ at the end of inflation.
Eq.  (\ref{6i}) suggests that at the end of inflation with $ \alpha H^2 \gg
m^2$ we will have perturbations ${\delta\rho_\sigma\over \rho_{\rm total}}
= | \delta _k^\rho | \, {\rho_\sigma\over \rho_{\rm total}}$ with  blue
spectrum,  $| \delta _k^\rho |\sim  \left({k\over H}\right)^{2\alpha/3}$.
Further evolution of these perturbations is strongly model-dependent. The main
point is that under certain conditions the factor ${\rho_\sigma\over \rho_{\rm
total}}$ may grow up to 1, whereas the shape of the potential may not change
considerably.  In what follows we will study one of these possibilities.  We
will skip the numerical coefficients of the order of one  since we are
interested only in order of magnitude estimates.

 According to \cite{book}, at the stage when the energy density is dominated by
${1\over 2}{M^2\phi^2 }$   the scale factor is  given by $a\simeq \exp (-2\pi
\phi ^2/M_p^2)$.
Consider the evolution of the modes of field $\sigma $   which
have the scales bigger than the horizon scale $H^{-1}$ during inflation.  The
comoving wavenumber $k$ for these modes satisfies the condition: $k\ll Ha. $
Neglecting the spatial derivatives in the equation for scalar field $\sigma
$ and assuming that during inflation $\alpha H^2 \gg m^2$ we obtain the
following equation for $\sigma _k$:  ~$3H\dot \sigma _k= - \alpha H^2 \sigma
_k$. The solution of this equation is $\sigma _k\propto a^{-\frac \alpha {3}}$.
 At the
moment of horizon crossing ($k\sim H_ka_k$) the amplitude of quantum
fluctuations is $\left| \sigma _k\right| ^2k^3\sim H_k^2$,  where $H_k$ and
$a_k$ are correspondingly the values of the Hubble constant and scale factor
taken at the moment when perturbation with the wavenumber $k$ crosses the
horizon.  This together with the relation $\sigma _k\propto a^{-\frac \alpha
{3}}$ yields the relation
\begin{equation}
\left| \sigma _k\right| ^2k^3\sim H_k^2\left( \frac {a_k}a\right) ^{\frac
{2\alpha} {3}}\sim ~H_k^2\left( \frac k{H_ka_k}\right) ^{\frac {2\alpha} {3}}
 \label{spectrum}
\end{equation}
for $k\ll Ha. $ Taking into account that $H_k\sim M\left( \ln (H_k/k)\right)
^{1/2}, $ we find that at the end of inflation
\begin{equation}
\left| \sigma _k\right| ^2k^3\sim M^2\left( \ln \frac{H_k}k\right)
^{1{-\frac \alpha {3}}}\left( \frac kM\right) ^{\frac {2\alpha} {3}}\mbox{
for }k\ll M.   \label{endspec}
\end{equation}
 Here we took the following normalization for the scale factor: $a=1$ at
the end of the inflation when   $\phi
\sim 1. $

Detailed evolution of these perturbations after inflation is model-dependent.
For simplicity  we assume here that the correction $\alpha H^2$ to the mass of
the field $\sigma$ disappears soon after the end of inflation. This is the case
for both of our models, but it may not be the case for supersymmetric models
studied in \cite{moduli}, where some corrections to our results will appear.
We will also assume that the particles $\sigma$ are either stable or decay very
late, and that the inflaton field decays into ultrarelativistic
particles with the energy density $\rho _r(t_f)\sim M^2$ immediately
after inflation.  For a long time these particles dominate. Hubble constant
decays as $H\propto M/a^2$
during radiation dominated epoch.  The modes with $ k\gg \max \left\{ ma,
Ha\right\} $ oscillate and their amplitudes decay as $a^{-1}$.  The
 modes with $k\ll \max \left\{ ma, Ha\right\} $ have practically
constant amplitude at the beginning and then start to oscillate with the
frequency $\sim m$ when the Hubble constant $H(a)$ drops below $m$.
Meanwhile their amplitudes decay as $a^{-3/2}$.   Using these facts one can
easily
calculate the spectrum of scalar field at $a>M/m$:
\begin{eqnarray}
\left| \sigma _k\right| ^2k^3 &\sim& a^{-3} M^2\left( \ln \frac{H_k}k\right)
^{1-{\frac {2\alpha} {3}}}\left( \frac kM\right) ^{\frac {2\alpha} {3}} F(k) \
, \nonumber\\
 F(k) &=&
\left\{\begin{tabular}{ll}
$\frac{M^2}{mk}$~~ & for $M>k>\sqrt {Mm}\ , $ \\
$\left(\frac Mm\right)^{3/2}$~~ & for $k <\sqrt {Mm} \ . $%
\end{tabular}
\right.   \label{op}
\end{eqnarray}

To make an estimate of  the energy density of fluctuations of the field
$\sigma$ one may  calculate   the
 potential'' energy density
\begin{equation}
\rho _\sigma \sim \langle\rho _\sigma ^{pot}\rangle=\frac
12m^2\left\langle \sigma ^2\right\rangle \ .   \label{pot}
\end{equation}
It is easy to check that the contributions of the other terms to the total
energy either the same order of magnitude or smaller than (\ref{pot}).  When $
H$ drops below $m$ term (\ref{pot}) correctly counts the leading
contribution to the energy density.  Taking into account that $\sigma _k$
depends only on $k\equiv \left| \stackrel{\rightarrow }{k}\right| $ and
integrating over the angles one can write (\ref{pot}) in the following
manner:
\begin{equation}
\left\langle \rho _\sigma \right\rangle \sim m^2\int  \left| \sigma _k\right|
^2k^3d(\ln
k)\ .   \label{en}
\end{equation}
Substituting (\ref{op}) in (\ref
{en}) and integrating over $k<aM$,  one obtains, for  $\alpha \ll 1, $
\begin{equation}
\left\langle \rho _\sigma \right\rangle \simeq \frac{M^{\frac
72}m^{\frac 12}}{a^3\alpha }\left( \frac mM\right) ^{\frac {\alpha}
{3}}\left(\ln \frac Mm+\frac{3 }\alpha \right) .   \label{ener}
\end{equation}

The produced $\sigma$-particles are not distributed homogeneously.  If the
scalar field is relatively stable and starts to dominate in some time after
inflation then their energy density fluctuations can become very relevant
for cosmology.  As we already mentioned, to estimate these fluctuations one
cannot simply apply the standard
methods developed in \cite{Pert} because in our case the   homogeneous
component of the field $\sigma$ vanishes.  A correlation function
characterizing
energy density fluctuations is
\begin{eqnarray}
\xi (r, t)=\frac 1{\left\langle \rho _\sigma \right\rangle ^2}\left(
\left\langle \rho _\sigma (\stackrel{\rightarrow }{x}, t)\rho
_\sigma (\stackrel{\rightarrow }{x}+\stackrel{\rightarrow }{r}%
, t)\right\rangle -\left\langle \rho _\sigma \right\rangle ^2\right) .
\label{corr}
\end{eqnarray}
The power
spectrum $\left| \delta _k^\rho \right| ^2$ is defined as follows:
\[
\xi \left( r, t\right) =\int \frac{dk}k\frac{\sin kr}{kr}\left| \delta
_k^\rho \right| ^2.
\]
To make an estimate of $| \delta _k^\rho |$ we will take into account
only the contribution of the potential term in the energy. After simple
calculations   one finds that
\begin{equation}
\left| \delta _k^\rho \right| ^2\sim \frac{m^4k^3}{\left\langle
\rho _\sigma \right\rangle ^2}\int \left| \sigma _{\stackrel{\rightarrow
}{k^{\prime }}}\right| ^2\left| \sigma _{\stackrel{\rightarrow
}{k}-\stackrel{\rightarrow }{k^{\prime }}}\right| ^2d^3 k \ .
\label{amp}
\end{equation}
Because we are
only interested   in large scale fluctuations which have been produced at
inflation, we can make a cut off at $k\sim  M$.  By substituting the spectrum
(\ref{op}) in  (\ref{amp}) and
taking into account (\ref{ener}) one finally gets:
\begin{equation}
\left| \delta _k^\rho \right| \sim C(k) \left( \frac {k}{M}\right) ^{\frac
{2\alpha} {3}}    \label{spe1}
\end{equation}
for $k\ll \min \bigl(M\left( \frac mM\right)^{1/2},~M\exp (-1 /2\alpha
)\bigr)$.  Here $C(k) = \sqrt{\alpha }\left(\ln \frac Mm+\frac{3 }\alpha
\right)^{-1}\left( \ln \frac Mk\right) ^{1-{\frac {\alpha} {3}}} \bigl({M\over
m}\bigr)^{\frac {2\alpha} {3}}$. Strictly speaking, the above expression for
$\left| \delta _k^\rho
\right| $ refers to the relative energy density fluctuations in $\left\langle
\rho _\sigma \right\rangle $ after inflation, at $a\gg M/m$,  but before
the field $\sigma $ starts to dominate.

 Let us suppose that the particles $\sigma$ at some stage began dominating the
energy density of the universe.  This does not necessarily imply that they
dominate the energy density at present, because they could decay later on.
However, as soon as they give the dominant contribution to density at some
moment,  perturbations of metric at all subsequent moments will be determined
by $| \delta _k^\rho  |$.  Here one should distinguish between two
possibilities. If the field $\sigma$ does not decay (as in the axion theory) or
decays in a hidden sector without changing the amount of photons in the MBR,
then the corresponding perturbations can be called isothermal, or entropy
perturbations.  However, if the field $\sigma$ dominates and decays
sufficiently early, then we essentially have a secondary reheating, and the
perturbations (\ref{6}) should be considered as   adiabatic perturbations with
blue spectrum. In a more general case by tuning the moment of decay one can get
a mixture of adiabatic and isothermal perturbations.

To evaluate possible implications of this result one should first of all
estimate the desirable value of the coefficient $\alpha$. Suppose   that, as
usual, perturbations of metric on the galaxy scale have the wavelength which
was about $10^{25}$ times greater than $H^{-1}\sim M^{-1}$ at the end of
inflation (our final result will not be very sensitive to this and the
subsequent assumptions). Suppose also that $C(k) \sim 10^2$ for $k\sim 10^{-25}
M$. Then the requirement that $\left| \delta _k^\rho \right| \sim 10^{-5}$ on
the galaxy scale gives us the estimate ${\frac {2\alpha} {3}} \sim 0.3$, which
corresponds to the spectral index $n \sim 1.6$. One can easily check that this
result is consistent with our assumption that $C(k) \sim 10^2$ for a broad
range of relations between $M$ and $m$. This means that an exact value of
$\alpha$ is not very much sensitive to our assumptions concerning other
parameters of our model. (We will specify the possible range of values of $M$
and $m$ in \cite{ML}.)  On the other hand, with ${\frac {2\alpha} {3}} \sim
0.3$ the amplitude of density perturbations falls more than $10$ times on the
way from the galaxy scale to the scale of horizon. This strongly suppresses the
microwave background anisotropy, thus resolving the problem of the  MBR
anisotropy produced by isothermal perturbations being 6 times greater than that
produced by adiabatic perturbations of the same magnitude. Of course it does
not mean automatically that this model can easily pass all  cosmological tests,
maybe in the end it will be desirable to consider combined effects produced by
adiabatic   and isothermal perturbations. This should be a topic of a separate
investigation. The main purpose of our work is to point out that there exists a
very simple but overlooked class of models where the perturbations of metric
have rather unusual properties: They are nongaussian, they can be either
isothermal or adiabatic,  and they have blue spectrum with $n > 1$.

Note that the nongaussian nature of   perturbations in these models is
especially interesting on scales smaller than $k_0^{-1}$. Indeed, on such
scales one can introduce an effective ``classical'' quasi-homogeneous field
$\sigma$. This field, however,  takes different values in different points. The
calculation of density perturbations in each small region of such type can be
performed in the standard way  \cite{Pert,muk1}, but the final result will
depend on our position in the universe \cite{Ax,ML}. This introduces an
additional type of the large-scale structure of the universe. For ${\frac
{2\alpha} {3}} \sim 0.3$ the corresponding scale $\sim k_0^{-1}$ is too small
to be of any real importance. It would be interesting, however, to study the
corresponding effects for $\alpha \ll 1$.

The models considered above can be easily generalized. For example, in
supersymmetric theories the terms proportional to $H^2$ appear not only in the
quadratic parts of the effective potential but to other parts as well.  A more
general toy model to study this effect is \cite{AD}
\begin{equation}\label{DRT}
V=- {1\over 2}(m^2 + {\alpha\over 2} H^2 ) \sigma^2 + {1 \over 4 M_p^2}
(\lambda m^2 + {\beta\over 2}  H^2) \sigma^4 \ .
\label{modulimodel}
\end{equation}
For $H \ll m_{\sigma}$,   the effective mass squared at the minimum of the
effective potential is   $2m^2$. Meanwhile in the early universe at $H \gg
m_{\sigma}$  the minimum is at $\sigma = \sqrt{\alpha/\beta}~M_p$, and the
effective mass squared of the field $\sigma$ at that stage is $\alpha H^2$. If
$\alpha \ll 1$, then at the stage of inflation  perturbations of the field
$\sigma$ are generated. The main difference between this model and the one
considered before is that in this model during inflation there exists a large
homogeneous   field $\sigma$. Therefore density perturbations can be calculated
in a more standard way \cite{Pert,muk1}. They are gaussian, with  blue spectrum
 growing at large $k$ as $k^{\alpha/3}$. In the   Affleck-Dine model of
baryogenesis   \cite{AD} this may lead to isothermal perturbations of baryon
density  with the spectral index   $n \approx 1+{2\alpha/3}$. We hope to return
to   this model and other issues discussed above in a separate publication
\cite{ML}.

The authors are grateful to R. Bond and L.A. Kofman for useful discussions. The
work of A.L. was  supported in part by  NSF grant PHY-9219345.
He is grateful to the Institute for Theoretical Physics, ETH,  Zurich for the
hospitality.
V.M. thanks  the Tomalla foundation for
financial support.


\begin{references}
\bibitem{Pert} V.F. Mukhanov and G.V. Chibisov, JETP Lett. {\bf 33},
532 (1981); V.~F.~Mukhanov and G.~V.~Chibisov, Sov. Phys. JETP {\bf 56}, 258
(1982);
S.W. Hawking, Phys. Lett. {\bf 115B},  295 (1982);
A.A. Starobinsky, Phys. Lett. {\bf 117B},  175 (1982);
A.H. Guth and S.-Y. Pi, Phys. Lett. {\bf 49},  1110 (1982);
J. Bardeen, P.J. Steinhardt and M. Turner, Phys. Rev. {\bf D28},  679
(1983); V.F. Mukhanov, JETP Lett. {\bf 41},  493 (1985).
\bibitem{muk1}  V.F. Mukhanov, H.A. Feldman, and R.H. Brandenberger, Phys.
Rept. {\bf 215}. 203 (1992).
\bibitem{Ax} A.D. Linde,     Phys. Lett. {\bf 158B}, 375 (1985);  L.A. Kofman,
Phys. Lett. {\bf 173B}, 400 (1986); L.A. Kofman and A.D. Linde, Nucl. Phys.
{\bf B282}, 555 (1987).
\bibitem{Liddle} A. Liddle and D. Lyth, Phys. Rep. {\bf 231}, 1 (1993).
\bibitem{ALyth} A.D. Linde and D.H. Lyth,   Phys.
Lett. {\bf B246},  353 (1990).
\bibitem{iso}
R. Stompor et al.,  astro-ph/9511087.
\bibitem {martin} M. White and J. Silk, astro-ph/9608177 (1996).
\bibitem{fluct}  T.S. Bunch and P.C.W. Davies, Proc. Roy. Soc. {\bf
A360},  117 (1978); A. Vilenkin and L. Ford, Phys. Rev. {\bf D26}, 1231 (1982);
A.D. Linde, Phys. Lett. {\bf 116B},  335 (1982);
A.A. Starobinsky, Phys. Lett. {\bf 117B},  175 (1982).
\bibitem{Lyth} D. H. Lyth, Phys. Rev. D {\bf 45}, 3394 (1992).
\bibitem{Schmid}  H.F. Muller and C. Schmid, gr-qc/9401020.
\bibitem{book} A. D. Linde, {\it Particle Physics and Inflationary
Cosmology} (Harwood, Chur, Switzerland, 1990).
\bibitem{moduli}  M. Dine, W. Fischler and D.
Nemeschansky, Phys. Lett. {\bf B136},  169 (1984);  G.D. Coughlan, R. Holman,
P.
Ramond and G.G. Ross, Phys. Lett. {\bf B140},  44 (1984); A.S. Goncharov, A.D.
Linde, and M.I. Vysotsky, Phys. Lett. {\bf 147B}, 279 (1984).
\bibitem{AD} M. Dine, L. Randall and S. Thomas,  Nucl. Phys. {\bf 458},  291
(1996);
G.W. Anderson, A.D. Linde, and  A. Riotto, Phys. Rev. Lett. {\bf 77},3716
(1996).
\bibitem{SS} M. Sasaki and B. Spokoiny, Mod. Phys. Lett.  {\bf A6}, 2935
(1991).
\bibitem{ML} V.F. Mukhanov and A.D. Linde, in preparation.

\end{references}
\end{document}